# Addressing NameNode Scalability Issue in Hadoop Distributed File System using Cache Approach

Debajyoti Mukhopadhyay
Department of Information Technology
Maharashtra Institute of Technology
Pune, India
debajyoti.mukhopadhyay@gmail.com

Chetan Agrawal
Department of Information Technology
Maharashtra Institute of Technology
Pune, India
cagrawal11@gmail.com

Devesh Maru
Department of Information Technology
Maharashtra Institute of Technology
Pune, India
deveshmaru16@gmail.com

Pooja Yedale
Department of Information Technology
Maharashtra Institute of Technology
Pune, India
poojareddy743@gmail.com

Pranav Gadekar
Department of Information Technology
Maharashtra Institute of Technology
Pune, India
ppgadekar.92@gmail.com

*Abstract*: **Hadoop is a distributed batch processing infrastructure which is currently being used for big data management. The foundation of Hadoop consists of Hadoop Distributed File System (HDFS). HDFS presents a client-server architecture comprised of a NameNode and many DataNodes. The NameNode stores the metadata for the DataNodes and DataNode stores application data. The NameNode holds file system metadata in memory, and thus the limit to the number of files in a file system is governed by the amount of memory on the NameNode. Thus when the memory on NameNode is full there is no further chance of increasing the cluster capacity. In this paper we have used the concept of cache memory for handling the issue of NameNode scalability. The focus of this paper is to highlight our approach that tries to enhance the current architecture and ensure that NameNode does not reach its threshold value soon.**

## I. INTRODUCTION

In the technology oriented age of today, there is a growing disparity between the amount of data being generated and the ability to process and analyze this data. Database management systems are designed to manage such large amount of data. The data being generated now a days from various sources such as social network, semantic Web, satellites, surveillance systems, streaming data and bioinformatics network is humongous in amount. Moreover, these datasets are highly unstructured and thus it is being very difficult to store and handle this data. These aspects of the datasets have made the existing database management systems (DBMS) a bit inadequate and inefficient to be deployed for management of the data being generated. This led to the birth of new database management systems which not only are capable but also are very much efficient in storing, querying, processing, analyzing and making the data useful in a better yet convenient way. Apache Hadoop is one such DBMS which provides a distributed processing of large data sets across clusters of computing. It is a distributed, highly scalable and fault tolerant in nature.

Hadoop was first developed as a Big Data processing system in 2006 at Yahoo! The idea is based on Google's MapReduce, which was first published by Google based on their proprietary MapReduce implementation. In the past few years, Hadoop has become a widely used platform and runtime environment for the deployment of Big Data applications [3][5]. The core of Hadoop includes filesystem, processing and computation resources and also it provides the basic services for building a cloud computing environment with commodity hardware. Hadoop provides a distributed file system and a framework for the analysis and transformation of very large data sets using the MapReduce paradigm which is the computation component of Hadoop. A Hadoop cluster scales computation capacity, storage capacity and IO bandwidth by simply adding commodity servers. Hadoop Distributed File System (HDFS) is the file system component of Hadoop where the data is stored [13].

HDFS has master/slave architecture. An HDFS cluster consists of a single NameNode, a master server that manages the file system namespace and regulates access to files by clients. In addition, there are a number of DataNodes, usually one per node in the cluster, which manage storage attached to the nodes that they run on. The NameNode keeps a reference to every file and block in the filesystem in memory, which means that on very large clusters with many files, memory becomes the limiting factor for scaling [7]. Thus when a NameNode memory is full it becomes difficult to just add another node in the cluster for further storage needs as the NameNode will not be able to handle this extra node and filesystem metadata that will be generated from it. This raises the NameNode scalability issue.

The goal of this paper is to provide a solution to reduce the use of NameNode memory space so that the problem of NameNode memory getting full which ultimately results in hindering the scalability of cluster will be delayed.

## II. LITERATURE SURVEY

The ever growing technology has resulted in the need for storing and processing excessively large amounts of data. The current volume of data is enormous and is expected to replicate over 650 times by end the year 2014, out of which, 85% would be unstructured. This is known as the 'Big Data problem'. Apache Hadoop is an open source project at Apache Software Foundation being built and used by a global community of contributors and users. Hadoop is a distributed batch processing infrastructure which is currently being used for big data management [3].

Hadoop is designed to be parallel and resilient. It redefines the way that data is managed and processed by leveraging the power of computing resources composed of commodity hardware [5]. And it can automatically recover from failures.

### A. Hadoop Architecture

The basic Hadoop architecture consists of two primary components viz. 1. Hadoop Distributed File System (HDFS) and 2. MapReduce

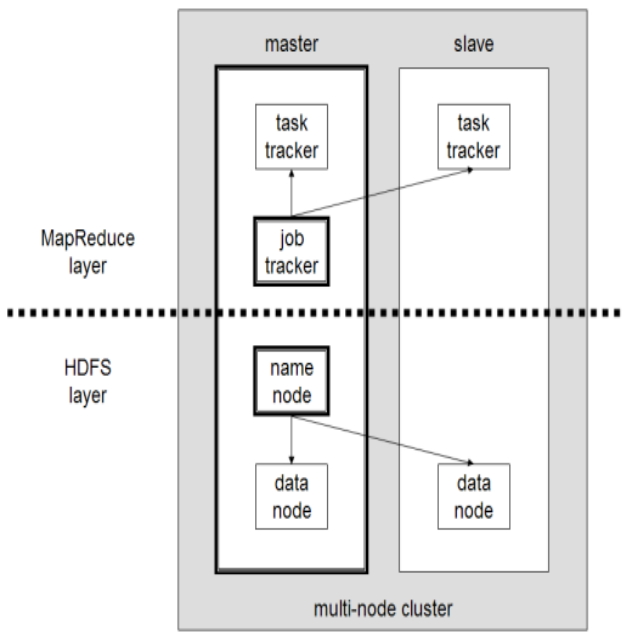

Fig. 1 Hadoop architecture

#### 1. Hadoop Distributed File System (HDFS layer)

Hadoop Distributed File System is a fault-tolerant distributed file system designed to run on "off-the-shelf" hardware. It has been optimized for streaming reads on large files whereas I/O throughput is favored over low latency. In addition, HDFS uses a simple model for data consistency where files can only be written to once [10]. HDFS assumes disk failure as an eventuality and uses a concept called block replication to replicate data across nodes in the cluster. HDFS uses a much larger block size when compared to desktop files systems. For example, the default block size for HDFS is 64 MB. Once a file has been placed into HDFS, the file is divided into one or more data blocks and is distributed to nodes in the cluster. In addition, copies of the data blocks are made, which again are distributed to nodes in the cluster to ensure high data availability in case of a disk failure. An HDFS cluster has two types of nodes operating in a master-worker pattern: a *NameNode* (the master) and a number of *DataNodes* (workers).

*a) NameNode*

The NameNode manages the filesystem namespace. It maintains the filesystem tree and the metadata for all the files and directories in the tree. This information is stored persistently on the local disk in the form of two files: the namespace image and the edit log [2]. The NameNode also knows the DataNodes on which all the blocks for a given file are located; however, it does not store block locations persistently, because this information is reconstructed from DataNodes when the system starts. A *client* accesses the file system on behalf of the user by communicating with the NameNode and DataNodes.

*b) DataNodes*

DataNodes are the workers of the file system. DataNode performs creation, deletion and copy of block under the NameNode's command. They store and retrieve blocks when they are told to (by clients or the NameNode), and they report back to the NameNode periodically with lists of blocks that they are storing. Each block replica on a DataNode is represented by two files in the local hosts native file system [2]. The first file contains the data itself and the second file is blocks metadata including checksums for the block data and the blocks *generation stamp*.

#### 2. MapReduce

MapReduce is a programming model for data processing. Hadoop can run MapReduce programs written in various languages. A MapReduce program is composed of a Map() procedure that performs filtering and a Reduce() procedure that performs a summary operation [8]. MapReduce works by breaking the processing into two phases: the map phase and the reduce phase. Each phase has key-value pairs as input and output, the types of which may be chosen by the programmer. The programmer also specifies two functions: the map function and the reduce function.

Here the application is divided into many small fragments of work, each of which can execute or re-execute on any node in the cluster. An important characteristic of Hadoop is the partitioning of data and computation across many (thousands) of hosts, and executing application computations in parallel close to their data [1] by using MapReduce paradigm.

In the nutshell in a typical MapReduce job, multiple map tasks on slave nodes are executed in parallel, generating results buffered on local machines. Once some or all of the map tasks have finished, the shuffle process begins, which aggregates the map task outputs by sorting and combining key-value pairs based on keys. Then, the shuffled data partitions are copied to reducer machine(s), most commonly, over the network. Then, reduce tasks will run on the shuffled data and generate final (or intermediate, if multiple consecutive MapReduce jobs are pipelined) results. When a job finishes, final results will reside in multiple files, depending on the number of reducers used in the job.

*B. NameNode memory management*

The NameNode stores its filesystem metadata on two important files viz. *fsimage* and *edits*. The *fsimage* contains a complete snapshot of the filesystem metadata whereas *edits* contains only incremental modifications made to the metadata [Hadoop ops]. On NameNode startup the *fsimage* file is loaded into RAM and any changes into edits are replayed, bringing the in-memory filesystem up to date.

*C. Hadoop and Hadoop Ecosystem*

Although Hadoop is best known for MapReduce and its distributed file system (HDFS, renamed from NDFS), the term is also used for a family of related projects that fall under the umbrella of infrastructure for distributed computing and large-scale data processing.

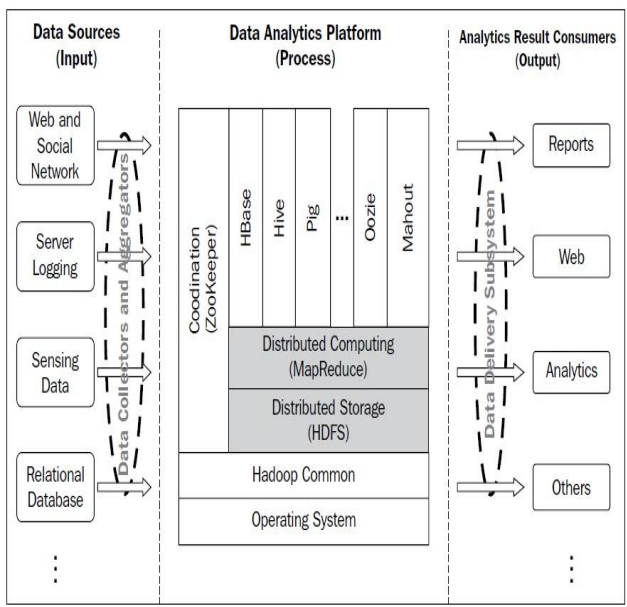

Fig. 2 Hadoop Ecosystem

A typical Hadoop-based Big Data platform includes the Hadoop Distributed File System (HDFS), the parallel computing framework (MapReduce), common utilities, a column-oriented data storage table (HBase), high-level data management systems (Pig and Hive), a Big Data analytics library (Mahout), a distributed coordination system (ZooKeeper), a workflow management module (Oozie), data transfer modules such as Sqoop, data aggregation modules such as Flume, and data serialization modules such as Avro [10].

*D. Problem with existing system*

The NameNode server in the Hadoop cluster keeps the track of filesystem metadata, it keeps a track of how your files are broken down into file blocks, which nodes store those blocks, and the overall health of the distributed filesystem. It maintains a catalog of all block location in the cluster which makes the NameNode the bookkeeper of HDFS.

Since the NameNode is a single container of the file system metadata, it naturally becomes a limiting factor for file system growth. In order to make metadata operations fast, the NameNode loads the whole namespace into its memory, and therefore the size of the namespace is limited by the amount of RAM available to the NameNode [9]. Thus when the memory available to the NameNode is full there are no chances for the cluster to grow limiting the number of active clients. Also a big HDFS installation with a NameNode operating in a large JVM where the memory space is almost full is vulnerable to frequent full garbage collections, which may take the NameNode out of service for several minutes [11]. Thus it is clear that memory available to the NameNode machine in a Hadoop cluster dictates the size of cluster and ultimately the number of active clients using a Hadoop based application.
.

III. PROPOSED WORK

*A. The basic idea of Cache memory*

A computer has a wide range of type, technology, performance and cost when it comes to complex memory management. Also a computer memory has a hierarchy from Central Processing Unit (CPU) registers to magnetic taped. Cache is one such memory type which comes second in the hierarchy which is an expensive yet fast access data storage device. A cache in generic terms is an intermediate buffer memory used to reduce the average time to access data. The cache is a smaller, faster memory which stores copies of the data from frequently used main memory locations. Cache memories are used in modern, medium and high-speed CPUs to hold temporarily those portions of the contents of main memory which are currently in use. Information located in cache memory may be accessed in much less time than that located in main memory. Thus, a central processing unit (CPU) with a cache memory needs to spend far less time waiting for instructions and operands to be fetched and/or stored. The instance when the main memory has to be accessed is only when data to be fetched is not in the cache i.e. a cache fault. Such cases arise very less number of times compared to cache hit.

*B. Design overview*

The basic concept of cache i.e. storing the frequently used data closer to the processor than the whole data and using it for faster operation can be applied in the NameNode memory management. Thus when the NameNode is operating in a heavy HDFS installation, the metadata records of frequently used data that is being used by most of the active clients can be kept in the NameNode memory space and the metadata records of least used data which sometimes can even be irrelevant can be kept at a different location and accessed as and when required. This way a lot of NameNode memory usage will be reduced which then can further be used to store more frequent data. Also with a less loaded HDFS NameNode the JVM garbage collection will be reduced which will avoid the problem of NameNode becoming irresponsive due to the excessive garbage collection.

*C. Architecture*

In this architecture, along with all the fundamental blocks of Hadoop one more commodity-hardware is attached. This added hardware is a considerably fast access memory device. It acts as the secondary storage device where the least recently used metadata will be stored. The metadata stored on it will be referenced only when the request has been made to access the data which is not being used frequently.

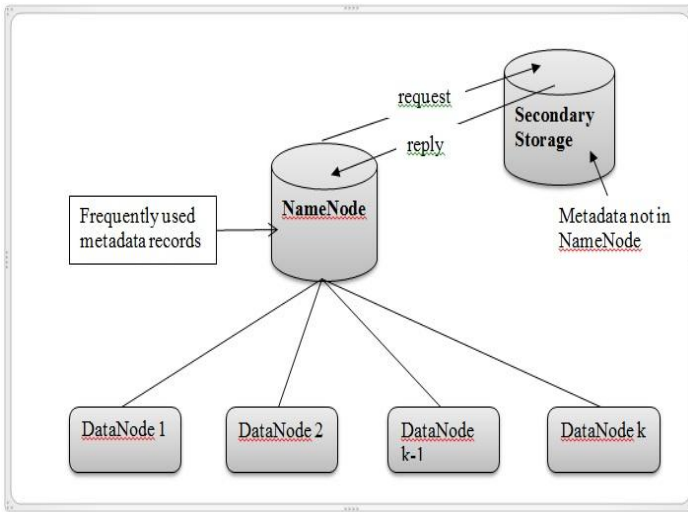

Figure 3

*1. Building Blocks of proposed architecture*

*NameNode:* The NameNode in the proposed architecture is same as that of it is in existing architecture except that it will not store the whole filesystem metadata. Instead the NameNode will store only those metadata records which are frequently and recently used. The NameNode will also implement a separation algorithm through which the least recently used metadata records will be separated out and will be stored on the secondary storage device.

*DataNodes:* The DataNodes will be same as that of existing system.

*Threshold Value:* A threshold value is the maximum amount of system memory space that the NameNode can use to store the Hadoop filesystem metadata. Memory usage reaching threshold value indicates that the system memory is about to be full and thus governs when the removal of metadata is to be carried out. We have installed Hadoop on a small cluster consisting of three machines. One dedicated machine as NameNode and two machines as DataNodes. The machine running NameNode has system memory i.e. RAM of 1GB available with it to load the Hadoop metadata.

Each metadata record requires approximately 600kiloBytes of memory. Thus 1GB of RAM will store approximately 1.8 million of metadata records. The threshold value for our implementation is the number of records the fsimage file contains as number of records directly indicates the memory requirements by NameNode.

As the fsimage file keeps on filling, the algorithm maintains a count on the number of metadata records in this fsimgae file contains. Once the number of records reaches a count of 1.26 million, indicating the memory requirement is about 700MB the separation algorithm is triggered to reduce the memory usage. The separation module separates out some of the records from the fsimage file based on the separation policy thus reducing RAM required to store those metadata records

*Separation Algorithm:* The separation algorithm will be used for moving the least recently used metadata from NameNode to secondary storage device. The separation algorithm takes into account two fields from the metadata records two decide whether or not to remove any metadata record. By default each metadata record keeps a track on last access time. We have added an extra field viz. count which indicated the frequency of use about each file into the cluster. Each time the access time is updated the count is increased by one indicating that the data/file represented by it is been used. Once triggered the separation algorithm works as follows:

1. Initially a mean of count which is the frequency of use of a file is calculated.
2. For each metadata record
    a. First check the last access time, if it indicates that the file has been accessed recently, then the metadata record is not removed irrespective of whether the file is used more frequently.
    b. If the access time indicates that the file is not been used for long time then the count i.e. the frequency of use of file is compared against the mean count that is calculated in the first step.
    c. If the count field in the metadata record is greater than the mean, suggesting that the data is used more frequently though it was not used recently. This metadata record is not removed.
    d. Rest of the metadata records are separated out and stored into another file viz. fsimage2.

This separation policy removes near about 30% of the metadata records from the fsimage file thus making space available for other purposes.

*Working:* In the proposed architecture, a threshold value will be defined on the NameNode memory space. As and when this threshold value will be reached the separation algorithm will run so as to remove the metadata records that are not being used. Initially the NameNode will keep storing the filesystem metadata as it comes until the predefined threshold is reached. Once the threshold will be reached the separation algorithm will come into play and remove some of the records based on the separation function. This separated data will be moved to secondary storage device

When the client performs a read operation the usual processing that is done by the NameNode to fetch metadata for requested file will be done and the client will be given back the file. The only difference will occur when the requested file is moved to the secondary storage as it was not recently used and the NameNode has reached threshold value. In this case the NameNode will not find the record in its memory and thus a request will be made to secondary device to fetch the metadata record and the metadata record will be removed from secondary device and will be loaded again on the RAM.

## IV. Implementation

*Experimental Setup:* Hadoop is deployed on three machines having Ubuntu 13.04 as linux version thus creating multimode cluster. One dedicated machine for NameNode and two for DataNodes. The NameNode stores the file system metadata into the file viz. fsimage and loads it each time NameNode comes alive. The data set used is a collection of 1.8 million text files collected from the weather survey. Files are added to NameNode and are accessed randomly leaving some of files untouched so that their frequency and last access time does not change. Once the count of metadata in fsimage reaches 1.2 million, the separation algorithm gets initiated. The algorithm removes the old metadata with low frequency of access and stores it on secondary storage. Initially, 1.2 million files were added to the cluster using put command. As the metadata for these files consumes about 700MB of RAM on NameNode, as we put 700 MB as our threshold for NameNode memory usage so the separation gets triggered and each time the metadata count in fsimage reaches 1.2 million i.e. 700 MB of NameNode memory usage the separation algorithm gets triggered and removes the not recently and frequently used data.

*Results:* As some of the files were left untouched and other were accessed randomly, the separation module remove approx. 0.3 Million files the first time and 0.25 million files for the second time and 0.28 million for the third time. This removal of the files has made available approximately 250 MB of RAM. Due to this free space, the Hadoop does not become unresponsive. The free space aids in faster operation by Hadoop NameNode as well as by local operating system.

A new file to store the separated metadata is created and the metadata that is separated is written to it. Due to less RAM consumption by NameNode, it becoming unresponsive and the requirement of restarting the cluster is avoided.

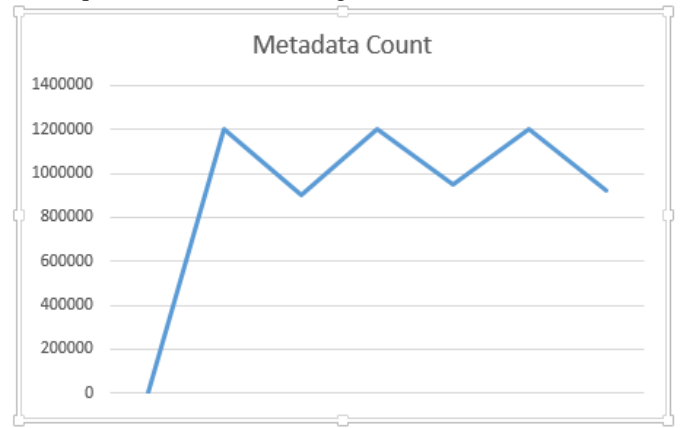

An observation that can be made is that every time our separation algorithm is triggered the data that remains in the main NameNode memory is in fact the metadata that has been recently added or has been frequently used and accessed. For future read operations this can be considered as a positive perspective, since whenever read request is made search operation carried on the NameNode will return the recent data.

## V. Conclusions

We have presented a way to reduce the NameNode memory consumption thus increasing the capacity of a Hadoop cluster. By implementing the cache concept in HDFS the two issues with Hadoop viz. NameNode being irresponsive due to garbage collection and cluster scalability issue can be solved. This way the Hadoop cluster will not reach the stage where the NameNode becomes irresponsive due to excessive JVM garbage collection as the HDFS will not be heavily loaded. Also as the NameNode will only store relatively more frequently used data the operations carried on the cluster will be faster and more efficient.